\documentclass[10pt]{article}

\usepackage{amsmath}

\usepackage{array}

\usepackage{appendix}

\usepackage{tocloft}                   

\usepackage{graphicx}

\usepackage{amsfonts}

\usepackage{amssymb}

\usepackage{mathrsfs}

\usepackage{yfonts}

\usepackage{euscript}

\usepackage{centernot}                 

\usepackage{ifsym}                     

\usepackage{lmodern}                   

\usepackage{upgreek}

\usepackage{mathtools}

\usepackage{color}

\usepackage{slantsc}
\usepackage{calligra}

\usepackage{bbold}          

\usepackage[T1]{fontenc}

\usepackage{epsf}

\usepackage{latexsym}

\usepackage{tipa}

\usepackage{makeidx}

\makeindex

\def\b{\begin{equation}}

\def\e{\begin{equation}}

\def\be{\begin{equation}}              

\def\ee{\end{equation}}

\def\beq{\begin{equation}}

\def\eeq{\end{equation}}

\def\bea{\begin{eqnarray}}

\def\eea{\end{eqnarray}}

\def\m{\mbox{ }}

\def\!{\hspace{-1.6667em}}

\def\n{\noindent}

\def\u{\underline}

\def\mH{\mbox{H}} 

\def\mJ{\mbox{J}}

\def\mS{\mbox{S}}                        

\def\bh{\u{\u{\mbox{h}}}  }            

\def\bA{\mbox{\bf A}}

\def\bQ{\mbox{\bf Q}}

\def\bh{\mbox{\bf h}}

\def\bupSigma{\mbox{\boldmath$\Sigma$}}                 

\def\cE{{\mathscr E}}

\def\sd{\mbox{\scriptsize d}}

\def\se{\mbox{\scriptsize e}}

\def\sf{\mbox{\scriptsize f}}

\def\sg{\mbox{\scriptsize g}}

\def\si{\mbox{\scriptsize i}}

\def\sm{\mbox{\scriptsize m}}

\def\sn{\mbox{\scriptsize n}}

\def\sbg{\mbox{{\bf \scriptsize g}}}

\def\sbg{\mbox{{\bf \scriptsize\sffamily g}}}

\def\sbcC{\mbox{\boldmath \scriptsize ${\cal C}$}}

\def\sbcG{\mbox{\boldmath \scriptsize ${\cal G}$}}

\def\sbcL{\mbox{\boldmath \scriptsize ${\cal L}$}}

\def\sumi2{\sum\mbox{}_{\mbox{}_{\mbox{\scriptsize $i$=1}}}^2}

\def\sumi3{\sum\mbox{}_{\mbox{}_{\mbox{\scriptsize $i$=1}}}^3}

\def\sumABcycles3{\sum\mbox{}_{\mbox{}_{\mbox{\scriptsize cycles $A,B$=1}}}^{3}}

\def\sumCDcycles3{\sum\mbox{}_{\mbox{}_{\mbox{\scriptsize cycles $C,D$=1}}}^{3}}

\def\sumj3{\sum\mbox{}_{\mbox{}_{\mbox{\scriptsize $j$=1}}}^3}

\def\sumk3{\sum\mbox{}_{\mbox{}_{\mbox{\scriptsize $k$=1}}}^3}

\def\prodiA1{\prod\mbox{}_{\mbox{}_{\mbox{\scriptsize $i$=1}}}^{A - 1}}

\def\d{\textrm{d}}                                                  

\def\pa{\partial}                                                   

\def\es{\m = \m}

\def\:={\m := \m}

\def\=:{\m =: \m}

\def\bLambda{\mbox{\boldmath$\Lambda$}}             

\def\sbiO{\mbox{\scriptsize\boldmath$O$}}

\def\nFrL{\mbox{\scriptsize$\mathfrak{L}$}}                         

\def\Hilb{\mbox{{\boldmath$\mathfrak{H}$}ilb}}                 

\def\scC{\mbox{\scriptsize ${\cal C}$}}                    
\def\scG{\mbox{\scriptsize ${\cal G}$}}                    

\def\scH{\mbox{\scriptsize ${\cal H}$}}                    

\def\scM{\mbox{\scriptsize ${\cal M}$}}                    

\def\Phase{\mbox{{\boldmath$\mathfrak{P}$}hase}}                     

\def\bFrR{\mbox{\boldmath$\mathfrak{R}$}}                            
\def\Rig-Phase{\bFrR\mbox{ig-}\Phase}                                
                                                                													   
\def\bphi{\mbox{\boldmath$\phi$}}                   %

\def\bFrR{\mbox{\boldmath$\mathfrak{R}$}}                            

\def\bFrR{\mbox{\boldmath$\mathfrak{R}$}}                            

\def\1mat{\u{\u{1}}}                                                 

\def\Positive-Modespace{\mbox{{\boldmath$\mathfrak{M}$}odespace$^+$}}

\def\POSITIVE-MODESPACE{\mbox{{\boldmath$\mathfrak{M}$}ODESPACE$^+$}}

\def\Kin-Hilb{\mbox{{\boldmath$\mathfrak{K}$}in-\Hilb}}                     

\def\Mid-Hilb{\mbox{{\boldmath$\mathfrak{M}$}id-\Hilb}}                     

\def\Dyn-Hilb{\mbox{{\boldmath$\mathfrak{D}$}yn-\Hilb}}                     

\def\5Star{\mbox{\Large$\star$}}              

\begin{document}

\begin{center}

\Large{\bf Comparative Theory of Background Independence}

\normalsize

\vspace{0.1in}

{\large \bf Edward Anderson}$^1$ 

\vspace{.1in} 

\end{center}

\begin{abstract}

The Lie claw digraph has recently been shown to control Background Independence and thus both the Problem of Time and the nature of Physical Law.    
This is established for Flat and Differential Geometry with varying amounts of extra mathematical structure. 
This Lie claw digraph has Generator Closure at its centre, 
                          Relationalism     at its root, 
			          and Assignment of Observables 
					  and Constructability from Less Structure Assumed (working if Deformation leads to Rigidity) on its other leaves.  
The centre is enabled by automorphisms and powered by the Lie Algorithm generalization of the Dirac Algorithm 
(itself holding for the canonical subcase, for which generators are constraints).  

\m 

\n We now explain how such claws are {\sl categorical} and thus {\sl universal} 
over all levels of mathematical structure that could be considered to be Background Independent.  
This follows from automorphisms both being categorical and supplying the centre with enough machinery to control the three peripheral aspects.  
Such claws in general thus merit a new name: UBIC (Universal Background Independence Claws) including allusion to their ubiquity.  
For a given level, the Problem of Time facets are ordered into two UBIC: the spacetime primality copy and the space/dynamics/canonical primality copy. 
These may or may not have a Wheelerian 2-way route connecting them: Constructability of Spacetime from Space and Intermediary Object Independence.    
The three cases of Constructability furthermore take one up and down levels like the ramps in a multi-storey car park. 
Wheeler matrices and ramp matrices (or more generally labelled digraphs) codify this information. 
These summarize the main features of each Background Independent Theory that can vary between levels,  
the Constructabilities and Intermediary Object Independence being selection principles rather than categorical. 
These selection principles indicate that Background Independence has reached maturity as a field of study.

\end{abstract}

$^1$ dr.e.anderson.maths.physics *at* protonmail.com

\section{Introduction}\label{Introduction}

Physical laws, in their usual mathematical setting of further-equipped Differential Geometry, 
are built upon Fig \ref{UBIC-Fig-1}.a)'s multi-aspected Lie structure \cite{XIV} 
%
{            \begin{figure}[!ht]
\centering
\includegraphics[width=0.85\textwidth]{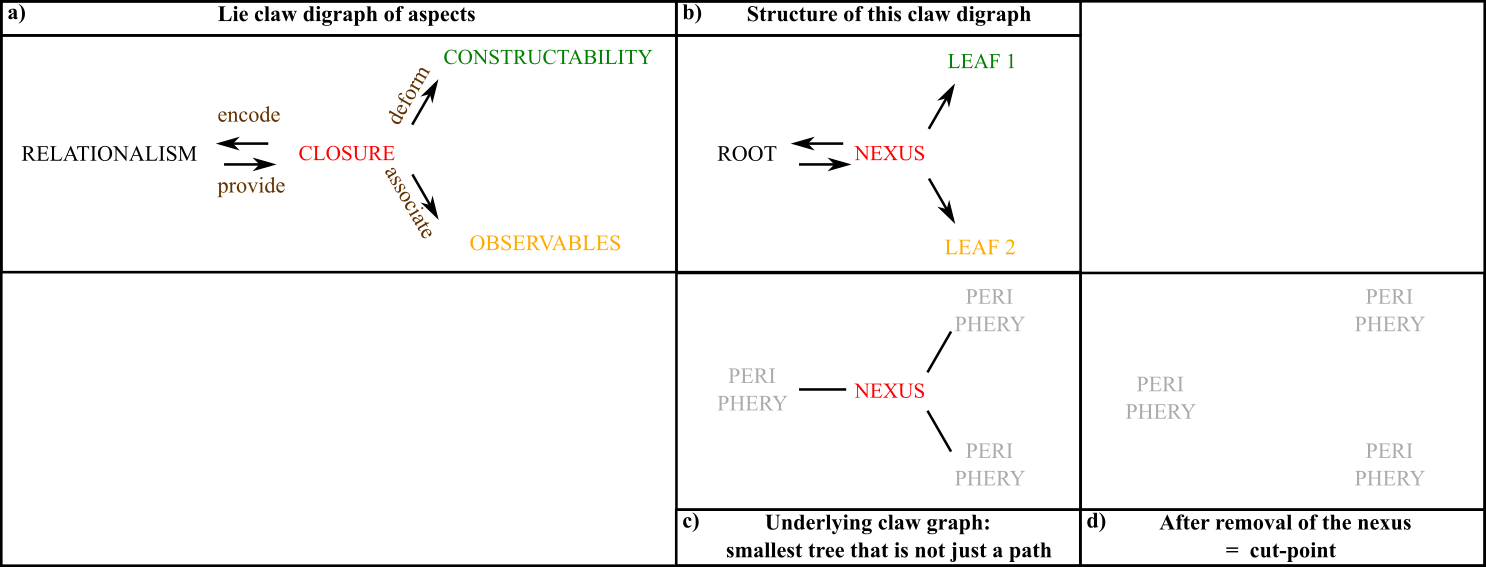} 
\caption[Text der im Bilderverzeichnis auftaucht]{\footnotesize{Lie Claw Digraph of Background Independence aspects a) 
with its vertices' graph-theoretic content in b), the underlying graph in c) and the effect of removing the nexus point in d): 
leaving the previously connected graph with 3 components.} } 
\label{UBIC-Fig-1} \end{figure}          }

\m 

\n This has the following four parts, corresponding to the four vertices (round brackets contain explicitly canonical counterparts and addenda). 

\m 

%
\n{\bf Lie aspect 0)} From its nexus,  
star-point,  
or cut-point \cite{Graph-1} 
status in the Lie claw digraph (Fig \ref{UBIC-Fig-1}.d), it is {\bf Closure} that is the central concept \cite{XIV} of Background Independence \cite{ABook, I, II, III, IV}.  
This means algebraic closure of generators $\sbcL$ (and other objects as detailed below) under Lie brackets $\mbox{\bf [} \m \mbox{\bf ,} \, \m \mbox{\bf ]}$.   

\m 

\n{\bf Lie strategy 0)} Closure is assessed by the Lie Algorithm \cite{XIV}.  
(In the canonical case, the Lie Algorithm specializes to the more familiar Dirac Algorithm \cite{Dirac, HT92, III} for constraint consistency, 
the generators now being constraints $\scC$.)

\m 

\n In more detail, first-class generators close under these while second-class ones do not \cite{XIV} 
(first- and second-class are more familiar in for the canonical setting's constraints).  
The Generalized Lie Algorithm \cite{XIV} permits the 6 following types of equation to arise from the generators $\sbcL$ 
(or constraints $\sbcC$, in which case one has the Dirac Algorithm \cite{Dirac, HT92, ABook}).  

\m 

\n i)   {\it New generators} $\sbcL^{\prime}$ (or constraints $\sbcC^{\prime}$) arising as integrabilities are reliably found thus. 

\m 

\n ii)  {\it Identities}:      equations reducing to $0 = 0$. 

\m 

\n iii) {\it Inconsistencies}: equations reducing to $0 = 1$. 
Including these incorporates \cite{XIV} an insight of Dirac's \cite{Dirac}, 
now promoted from its more restricted context of Poisson brackets algebras of classical constraints to the generalized Lie context.   
The Generalized Lie Algorithm thereby gains the capacity to reject candidate theories' sets of generators.  

\m 

\n iv)  {\it Rebracketing} using `{\it Lie--Dirac brackets}' in the event of encountering {\it second-class objects} 
(generalizing use of Dirac brackets \cite{Dirac, HT92} to eliminate second-class constraints).  

\m 

\n v)  `{\it Specifier equations}' are also possible in the presence of an appending process. 
(Dirac's \cite{Dirac} appending of constraints to Hamiltonians $H$ using Lagrange multipliers $\bLambda$:  

\n\be 
H \longrightarrow H + \bLambda \cdot \sbcC \m . \m )
\ee 
These specify which forms a priori free appending variables take. 

\m 

\n vi) {\it Topological obstruction terms} \cite{Dirac, ABook} such as anomalies [though the current Article just proceeds locally].   

\m 

\n The Generalized Lie Algorithm {\it terminates} if \cite{III, XIV} one of the following occurs. 

\m 

\n 0)   It {\it hits an inconsistency}, 

\m 

\n I)   It {\it cascades to inconsistency}. 

\m 
         												                
\n II)  It {\it cascades to triviality}.  
	
\m
	
\n III) It {\it arrives at an iteration producing no new objects} while retaining some degrees of freedom.   

\m 
																						
\n Successful candidate theories terminate by III), producing Lie algebraic structures of generators $\nFrL$ (or of first-class constraints).   

\m 

%
\n{\bf Lie aspect 1)} These generators can be viewed as provided by {\bf Relationalism}'s `root' in the Lie claw digraph. 
In some approaches \cite{ABook}, one alternatively starts with Closure and then encodes Relationalism. 

\m 

\n{\bf Lie strategy 1)} Physically meaningless transformations are incorporated by means of Lie derivatives \cite{ABook, I, II, V, VI, X}.

\n\be 
\pounds_{\u{X}} \m . 
\ee 
In some approaches \cite{ABook, VII}, one cycles between Relationalism and Closure until a Relationalism is found whose closure is guaranteed by the Lie Algorithm. 

\m 

\n In the dynamical setting, Relationalism splits into \cite{APoT2, ABook, I, II, V, VI} 
Temporal Relationalism and Configurational Relationalism (spatial and instantaneous-internal).

\m 

\n Configurational and Spacetime Relationalisms are implemented by Lie corrections with respect to physically-irrelevant transformations in a fairly obvious manner.
For GR, on the one hand Configurational Relationalism with respect to spatial diffeomorphisms provides the momentum constraint $\u{\scM}$, 
On the other hand, Spacetime Relationalism with respect to spacetime diffeomorphisms involves 
e.g.\ the point identification map from the geometrical study of cosmological perturbations \cite{Stewart}.

\m 

\n{\bf Temporal Relationalism Strategy 1)} 
Temporal Relationalism is implemented by making no use of extraneous times, extraneous time-like variables, or label times.  
The last of these requires, firstly, use of configuration--change variables $\bQ$, $\d \bQ$.
Since $\d = \pounds_{\sd}$: deparametrized version of $\pa/\pa \lambda = \pounds_{\pa/\pa \lambda}$, this continues to be under the remit of use of Lie derivatives.
Secondly, implementation by the deparametrized version of a reparametrization-invariant action: 

\n\be 
{\cal S} = \int \d(\mJ\mS(\bQ, \d\bQ)  \m \mbox{ homogeneous linear in }  \m  \d \bQ  \m ,    
\ee 
$\d(\mJ\mS)$ standing for {\it Jacobi--Synge arc element} \cite{Lanczos}.
An argument of Dirac \cite{Dirac} then applies, by which at least one primary constraint must arise. 
This gives Temporal Relationalism's way of encoding GR's Hamiltonian constraint, $\scH$ \cite{BSW, RWR, FileR, ABook} 
(the more usual way -- variation with respect to the lapse \cite{ADM} -- being barred by lapse being a time-like variable \cite{ABook}).    
Thirdly, subsequent work is to stay within \cite{ABook} a Principles of Dynamics that involves changes rather than velocities among its variables 
\cite{ABook, FileR, AM13, I, V, VI, VII, VIII}. 
(Momenta and Poisson brackets, however, remain licit in such a formulation, 
by which the Hamiltonian formulation and passage to the quantum remain largely unaltered by Temporal Relationalism.)   

\m 

\n{\bf Temporal Relationalism strategy 1$^{\prime}$)} Finally, a notion of time is to be recovered at an emergent level in the following `Machian' manner.
In dynamical or split space-time approaches, time is to be abstracted from change,  

\n\be 
t^{\se\sm} = t^{\se\sm}(\bQ, \d \bQ)  \m .  
\ee   
See \cite{ABook, I, FileR} and references therein for details of how this expression is obtained 
by viewing the corresponding primary constraint as `an equation of time'. 
Beyond this point one can use $\bQ(t^{em})$, alongside standard formulations of the Principles of Dynamics.

\m 

%
\n{\bf Lie aspect 2)} Given phase space or the space of spacetimes, observables $\sbiO$ are the {\bf associated functions thereover}. 
In the presence of generators (or constraints), restricted (constrained) observables are such functions that additionally brackets-commute 
\cite{DiracObs, HT92, K92-I93, K93, AObs, ABook, III, VIII} with these.  
I.e.\ 

\n\be 
\mbox{\bf [} \, \sbcL, \, \sbiO \, \mbox{\bf ]}  \m = \m 0 \m \mbox{ or } \m  \approx 0 \m \m :
\ee 
Dirac's notion \cite{Dirac} of weak equality extended to Lie Theory. 

\m 

\n The Jacobi identity moreover dictates \cite{AObs} that these notions only make sense after Closure has been ascertained (hence the downward arrow in Fig 1.c), 
and that these observables themselves close to form algebraic structures.  

\m  

\n Our zero brackets condition moreover translates to a first-order flow PDE system \cite{Lee2} 
amenable to (a slight extension of) Lie's Integral Approach to Geometrical Invariants \cite{Lie, Olver2}.  

\m 

\n{\bf Lie strategy 2) Integral Approach to Invariants} by solving first-order differential equation systems 
                                                        obtained by writing out Lie brackets of generators with zero commutants.   

\m 

%
\n{\bf Lie aspect 3)} Once Closure is attained, the generators $\sbcL$ can, on the other hand, be Deformed \cite{G64-NR66}  

\n\be   
\sbcL \m \longrightarrow \m \sbcL_{\alpha}  \es  \sbcL + \alpha \, \bphi
\ee 
for parameter $\alpha$ and functions $\bphi$, to see if Lie Algorithm consistency resists such alterations. 

\m 

\n{\bf Lie strategy 3)} Mathematical capacity for this to occur \cite{Higher-Lie, XIV} is conferred by {\bf Rigidity} \cite{G64-NR66} 
of the underlying undeformed generator algebraic structure. 
This can allow us to reach the same conclusion under assumptions of less structure.
We term this {\bf Constructability} \cite{IX} (the actual name of aspect 3).
In particular, GR is Rigid \cite{AM13}.
Rigidity is in turn underlied by cohomological conditions 

\n\be 
\mH^2(\nFrL, \, \nFrL) = 0  \m ,  
\ee 
and can be taken \cite{XIV} to provide a Selection Principle in the Comparative Theory of Background Independence (CoToBI). 
The Lie Algorithm can moreover {\bf branch} (\cite{HT92} say `bifurcate') 
corresponding to setting each of a string of multiplicative factors to zero giving a distinct consistent possibility. 
Schematically, 

\n\be
\mbox{\bf [} \sbcL \mbox{\bf ,} \, \sbcL \mbox{\bf ]}  \es  A \times B \times \mbox{(trivializing or inconsistent)}  \m , 
\label{Branch}
\ee 
so we strongly set $A = 0$ (branch 1) or $B = 0$ (branch 2) to avoid the (trivializing or inconsistent) branch.

\m 

\n On the one hand, the Dirac Algorithm subcase of inconsistencies arising under Deformation is better known; see \cite{AM13, A-Brackets} and references therein. 
On the other hand, the Lie case's Deformations and Rigidities -- if not assessment of inconsistencies -- was done much earlier in a different literature 
\cite{G64-NR66}.  
That the Generalized Lie Algorithm has the capacity to pull this off in cases other than the Dirac Algorithm 
is exemplified by provision of new foundations for Flat Geometry \cite{A-Brackets, XIV}. 
I.e.\ the two alternative `top geometries' here -- Conformal versus Projective -- arise as branching in a Deformation and Rigidity analysis.  
This occurs both for Space from Less Space Structure assumed and for its indefinite flat spacetime counterpart.  

\m  

\n Modern paradigms of Physics usually involve more than one realization of the Lie Claw Digraph. 
In particular, most involve both of the following. 

\m 

\n I)  {\bf Spacetime primality}. 

\m 

\n II) {\bf Spatial/dynamical/canonical primality}. 

\m 

\n In the latter, one of phase space, configuration space or some other half-polarization of phase space usually play a leading role.  
The former's counterpart of this involves some space of spacetimes.  

\m 

\n There can moreover be Wheelerian 2-way routes \cite{MTW} between realizations A) and B), which we pose as follows.    

\m 

\n {\bf Route A)} {\bf Constructability of Spacetime from Space} \cite{RWR, AM13, ABook} and 

\m 

\n {\bf Route B)} {\bf Foliation Independence} \cite{ABook} (of spacetime into spatial slices).  
%
{            \begin{figure}[!ht]
\centering
\includegraphics[width=0.85\textwidth]{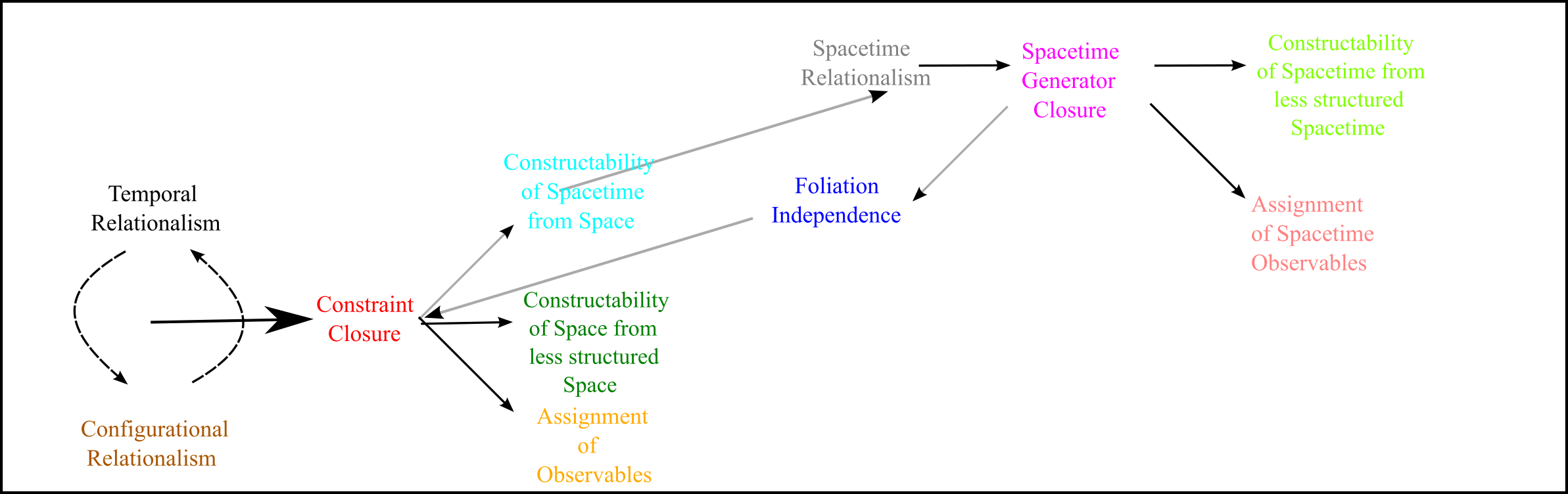} 
\caption[Text der im Bilderverzeichnis auftaucht]{\footnotesize{Classical order of incorporation of Background Independence aspects.
I.e.\ of overcoming corresponding Problem of Time facets, consisting of two copies of the Lie claw (black arrows) 
                                                        and a Wheelerian two-way route between them (grey arrows).} } 
\label{New-Gates} \end{figure}          }

\m 

\n{\bf Strategy A)} Like for all Constructabilities, Spacetime Constructability's success is contingent on encountering {\bf Rigidity} 
upon applying Deformation to the canonical constraints \cite{XIV}.  	  
	  
\m 

\n{\bf Strategy B)} {\bf Refoliation Invariance} (as outlined in Fig 3.a-b) resolves Foliation Independence in the case of GR.

\m 

\n We thus arrive at Fig 2's realization of Background Independence Aspects \cite{XIV}. 
This resolves (flat, curved or differential) geometrical levels of structure's manifestation of the Problem of Time 
\cite{Battelle-DeWitt67, K92-I93, APoT, APoT2, ABook, I, II, III, IV}.
Note that, for now, this is just classical, local and with no claims on being unique, 
which we summarize by terming it the classical ALRoPoT (A Local Resolution of the Problem of Time) \cite{ALett, I}.
The Theory of Background Independence our resolution employs is in turn classical ALToBI (A Local Theory of Background Independence) \cite{I, A-CBI}.    

\m

\n The names of Problem of Time facets, and Background Independence aspects are further explained in Appendix A.  
We moreover generalize away from additionally equipped Differential Geometry to 
{\sl every other} level of structure which could be considered to be Background Independent \cite{ASoS, ABook} in Secs \ref{Cats} (the claws) 
                                                                                                             and \ref{Wheelerian} (the Wheelerian 2-way routes).
Sec \ref{TD} considers towers of (almost-)copies of Fig \ref{New-Gates}, which we further characterize using `Wheeler' and `ramp' matrices.  
More generally, branching and looping occurs, by which {\sl digraphs} of levels of mathematical structure are required.  
We then characterize the relevant CoToBI information by `Wheeler' and `ramp' multilabelled digraphs.   
While this research goes further than most previous considerations of Background Independence, 
we remind the reader that a subset of it -- `Topology Change' Background Independence -- has a 30-year history \cite{W88, I89-91, GH92, ASoS, ABook, A-Leibniz}).  
While the Constructabilities and Reallocation of Intermediate-Object Invariance can categorically be {\sl posed}, 
they are not always expected to be realized. 
These are thus Selection Principles \cite{Higher-Lie, XIV} for an {\sl actual functioning} CoToBI.   
It is moreover these aspects that are summarized by the `Wheeler' and `ramp' objects (matrices, or, more generally, labelled digraphs),   
by which these are expected to be crucial toward most future studies concerning Background Independence.

\vspace{10in}

\section{Universal Background Independence Claw}\label{Cats}

The Lie claw's four aspects turn out to already be categorical in name (Fig \ref{UBIC-App-Fig}). 
For these                              to be categorical by nature as well, we rephrase them as follows. 

\m 

\n{\bf Categorical aspect 0)} {\bf Closure}, now referring to algebraic closure of generators $\sbcG$ (etc.\ as detailed below) under general brackets 

\n\be
\mbox{\bf |[} \m \mbox{\bf ,} \, \m \mbox{\bf ]|}_{\sbg}  \m , 
\ee   
remains the central aspect (this bracket could be $n$-ary).  

\m 

\n{\bf Categorical strategy 0)} Closure is to be assessed by the {\bf General Brackets Algorithm}, which takes the below form.  

\m 

\n The General Brackets Algorithm permits the 6 following types of equation to arise from an incipient set of generators $\sbcG$.  

\m 

\n i)   {\it New generators} $\sbcG^{\prime}$ arising as integrabilities are reliably found thus. 

\m 

\n ii)  {\it Identities}:      equations reducing to $0 = 0$. 

\m 

\n iii) {\it Inconsistencies}: equations reducing to $0 = 1$. 
Including these further generalizes the scope of Dirac's insight \cite{Dirac}, now to the general brackets algebraic structure context.   
The General Brackets Algorithm thereby gains the capacity to reject candidate theories' sets of generators, thus serving as a Selection Principle.  

\m 

\n Let us further extend Dirac's notion of first- and second-class constraints to the following. 

\m

\n{\bf Definition 1} {\it General-brackets-first-class  generators} are those that close under our general brackets, 
                 and {\it general-brackets-second-class generators} are ones which do not.  

\m 

\n iv)  {\it Rebracketing} uses `{\it general--Dirac brackets}' in the event of encountering {\it general-brackets-second-class generators}.  
This further generalizes Dirac brackets \cite{Dirac, HT92}, now in the context of general brackets algebraic structures.  

\m

\n v) `{\it Specifier equations}' are also possible in the presence of an appending process,  

\n\be
\cE \longrightarrow \cE + \bA \cdot \sbcG   \m ,   
\ee
for $\cE$ an encoding function (e.g.\ a Hamiltonian) and $\bA$ auxiliary variables used to append generators $\sbcG$.  

\m 

\n vi) {\it Topological obstruction terms} such as (general brackets algebras' generalizations of) anomalies.  

\m 

\n The General Brackets Algorithm {\it terminates} if \cite{III, XIV} one of I) to IV) (Sec \ref{Introduction}) occurs 
(those are already categorical in name and nature). 

\m 

\n{\bf Remark 0.1} Nijenhuis and Nambu examples of Brackets Algorithms have recently been given \cite{Nijenhuis, Nambu}.  

\m

\n{\bf Remark 0.2} Persistence of the claws away from Lie's domain is mostly for the following reason. 
Given a level of structure $L$, its automorphism group $Aut(L)$ is categorically well-defined. 
This being a group (or a generalization thereof), it has a product structure by which Closure can be assessed.  
Closure's centrality then turns out to be enough for the whole claw to categorically work out. 

\m 

\n That Temporal Relationalism can provide a further non-automorphism source of generators 
is mitigated by brackets algebraic structures also being categorically meaningful.  

\m

\n Let us end by noting the good track records of automorphisms and brackets algebraic structures as regards providing powerful and innovative Mathematics, 
with e.g.\ automorphisms underlying e.g.\ the Kleinian approach to Geometry, approaches to Topology in parallel to that, and Galois Theory.  

\vspace{10in}  

\n{\bf Categorical aspect 1)} The above generators can be viewed as provided by {\bf Relationalism}'s `root'; 
conversely, one could start with Closure and then encode Relationalism. 

\m 

\n{\bf Categorical strategy 1)} Physically meaningless transformations are incorporated by means of drag (generalized Lie) derivatives-or-differences,  

\n\be 
\pounds^{\sg}_{\u{X}} \m .  
\ee 
\n{\bf Remark 1.1} As an example of a nontrivially generalized Lie derivative, see e.g.\ \cite{Nijenhuis} for the Schouten--Nijenhuis Lie derivative.  

\m

\n{\bf Remark 1.2} The Temporal to Configurational Relationalism distinction is not however necessarily categorical.
For instance, the two cannot be split in e.g.\ Supergravity \cite{ABook}. 
%
  
\m 

\n{\bf Remark 1.3} The root comes with its own caveat: we need the categories in question to be {\sl dynamical}. 
An incipient way to view this is as not involving generalized configurations $\bQ$ but configurations as functions of time $\bQ(t)$. 
This survives reformulation under Temporal Relationalism, via $\bQ(\lambda)$ -- mere label time dependence -- 
to $\bQ, \d \bQ$ (continuum) or $\bQ, \Delta \bQ$ (discrete) configuration--change variables.
This done, we can again formulate a deparametrized version of reparametrization-invariant actions elements over configuration-change space 
(categorical version of Temporal Relationalism Strategy 1).  

\m 

\n{\bf Remark 1.4} It is useful here to clarify that cases involving $\Delta \bQ$ do moreover retain a discretized variant of the Calculus of Variations, 
e.g.\ along the lines of Dittrich and Hoehn \cite{DH13}.  
Differences in place of derivatives is a more basic and well-known consideration (extending to cover the above mention of `Lie differences').  
In this way, everything else used in the claw remains poseable, at least, at the categorical level.  

\m 

\n{\bf Remark 1.5} The categorical version of Temporal Relationalism Strategy 1 is, finally, 
that in space/dynamical/canonical approaches, time is to be abstracted from differential-or-discrete change.  
    
\m
  
\n{\bf Categorical aspect 2): Assignment of Observables} Given generalized phase space or the generalized space of spacetimes, 
observables $\sbiO$ are the {\it associated functions thereover}. 
In the presence of generators, restricted observables are such functions that additionally general-brackets-commute 
\cite{DiracObs, HT92, K92-I93, K93, AObs, ABook, III, VIII} with these.  
I.e.\ 

\n\be 
\mbox{\bf |[} \, \sbcG, \, \sbiO \, \mbox{\bf ]|}_{\sbg}  \m = \m  0 \m \mbox{ or } \m \approx \m 0  \m :
\ee 
Dirac's notion \cite{Dirac} of weak equality extended to general brackets algebraic structures. 
Given a general brackets algebraic structure, asking for its zero commutants is categorically meaningful as well.  

\m 

\n{\bf Remark 2.1} The bracket in question could be $n$-ary; 
it then makes no difference how many slots involve $\scG$'s leaving the other slots filled with $\sbiO$'s \cite{Nambu}.  

\m  

\n{\bf Remark 2.2} Our zero brackets condition moreover translates to a first-order flow differential-or-difference equation system.   
Field Theory has a rather less explored FDE counterpart to this \cite{VIII}.

\m 

\n{\bf Categorical strategy 2)} {\bf Integral-or-Sum Approach to Invariants}, 
                                by solving first-order differential-or-difference equation systems 
                                obtained by writing out general brackets of generators with zero commutants.   
This clearly extends Lie's Integral Approach to Geometrical Invariants.  

\m 

\n{\bf Facet interference 1-2}  At the Differential Geometry level of structure, it is the Jacobi identity that dictates the following.

\m 

\n i)  That Assignment of Observables only makes sense after Closure is ascertained [the downward arrow in Fig 1.c)]. 

\m

\n ii) That these observables themselves close to form algebraic structures.  

\m

\n It is then of interest whether these observations survive generalization away from Lie algebras. 
\cite{Nambu} shows that Filippov's generalization of the Jacobi identity in the Nambu brackets setting retains these properties. 
It is not however yet clear whether the other main way of generalizing Lie brackets -- to Loday brackets \cite{KS} -- retains these properties.\footnote{Be that 
in general or within some subcase, the Vinogradov brackets mentioned in \cite{Nijenhuis, Nambu} being one such subcase of theoretical interest.}

\vspace{10in}

\n{\bf Categorical aspect 3) Constructability}. Once Closure is attained, the generators $\sbcG$ can, on the other hand, be Deformed \cite{G64-NR66}  

\n\be   
\sbcG \m \longrightarrow \m \sbcG_{\alpha}  \es  \sbcG + \alpha \, \bphi
\ee 
for parameter $\alpha$ and functions $\bphi$, to see if Brackets Closure Algorithm consistency can resists such alterations. 
This can permit some levels of structure to be reached under a priori assumptions of less structure, hence the name `Constructability'. 

\m 

\n{\bf Categorical strategy 3)} Mathematical capacity for this to occur \cite{Higher-Lie, XIV} 
is conferred by {\bf Rigidity} \cite{G64-NR66} of the underlying undeformed generator algebraic structure. 

\m

\n{\bf Remark 3.1} Given a general brackets algebraic structure, its deformations are also categorically meaningful. 

\m 

\n{\bf Remark 3.2} Rigidity is in turn underlied by cohomological conditions. 
These can differ in the following ways. 

\m

\n i) how many slots they have (the Introduction's has two).

\m

\n ii) They might also differ as to the order taken by the diagnostically-crucial cohomology groups (the Introduction's is a second cohomology group).   
 
\m  
 
\n{\bf Remark 3.3} Rigidity occurring can be taken \cite{XIV} to provide a Selection Principle in CoToBI. 

\m 

\n{\bf Remark 3.4} The General Brackets Algorithm can moreover {\bf branch},
corresponding to setting each of a string of multiplicative factors to zero giving a distinct consistent possibility. 

\m 

\n{\bf Remark 3.5} How general an algebraic structure do we presently know how to deform?  
Lie algebroids, Nambu, Nijenhuis and Gerstenhaber algebras having already been surveyed in \cite{Nambu}, 
I now point to recent revival \cite{D19} of work on deformations of Loday algebras \cite{B97}.  

\m 

\n{\bf End-Remark I} do not know much yet about extent of literature available for the difference versions of Assignment of Observables 
or Constructability by Deformations encountering Rigidity.  

\m

\n{\bf End-Remark II} It is not however suitable to continue to name these claws after Lie, since they have transcended the realm of equipped Differential Geometry. 
We instead coin the name {\bf UBIC} -- {\bf Universal Background Independence Claw} -- 
these claws moreover indeed being ubiquitous in the level-by-level study of Background Independence.

\section{Wheelerian 2-way Routes in general}\label{Wheelerian}

\n{\bf Categorical aspect A) Constructability of Spacetime from Space}.   

\m 

\n{\bf Remark A} Level by level losses in distinction between spacetime and space are documented in \cite{ABook}.  
For instance, causality and signature drop out still with the Differential Geometry family of levels of structure.  
`Bigger' versus `smaller' distinctions between the two persists,
such as poset antichains (space) versus entire posets (spacetime), or as a subset within a larger unequipped set.
 
\m  

\n{\bf Categorical strategy A)} is {\bf Rigidity}, as per all Constructabilities in the current Article.  

\m 
 
\n{\bf Remark B.1} Foliation Independence \cite{K92-I93}, and its resolution by Refoliation Invariance (\cite{T73} and \ref{Pentagon}.a-b), 
also require categorical generalization.   

\m 

\n{\bf Categorical Aspect B) Intermediary-Object Independence}

\m 

\n{\bf Categorical Strategy B)} Reallocation of Intermediary-Object (RIO) Invariance. 
This retains the algebraic commuting-pentagon structure visible in Fig \ref{Pentagon}.c).  

\m 

\n{\bf Remark B.2} In the context of a smaller structure (space) within a bigger structure (spacetime), one can refer to these as `Slicing Independence' 
                                                                                                                              and `Reslicing Invariance'.

\m 

\n{\bf End-Remark III} The three Constructabilities and RIO Invariance are to be Selection Principles in CoToBI \cite{A-CBI}.    
%
{            \begin{figure}[!ht]
\centering
\includegraphics[width=1.0\textwidth]{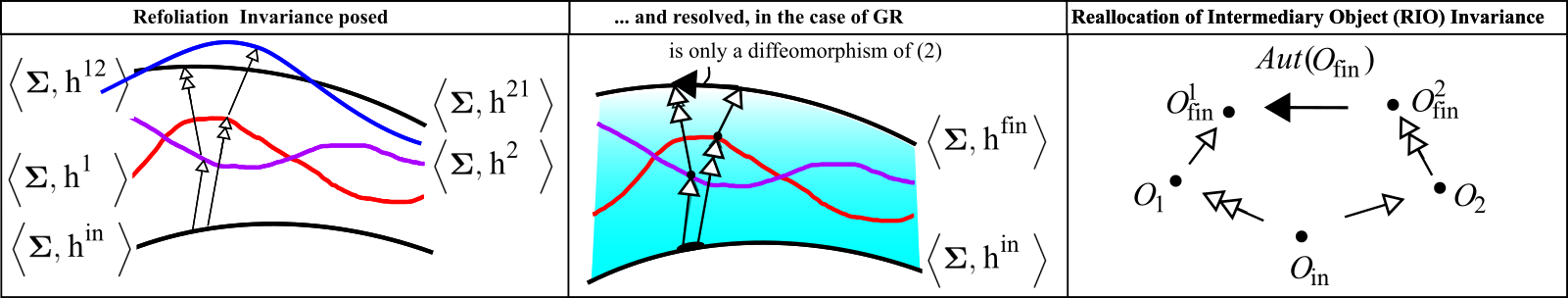}
\caption[Text der im Bilderverzeichnis auftaucht]{ \footnotesize{a) Poses and b) resolves GR's Refoliation Invariance.
I.e.\ whether in going from an initial object to a final object, proceeding (= applying the GR Hamiltonian constraint $\scH$) via the red    surface 1 
                                                                                                                               or the purple surface 2 
                                                                 causes discrepancy by at most a diffeomorphism of the thus-shared final surface, fin.
$\bupSigma$ is here a fixed spatial topology, whereas the $\bh$ are spatial metrics.  																 
Commuting pentagon from $O_{\si\sn}$ to $O_{\sf\si\sn}$ via two distinct allocations of intermediary objects, $O_1$ and $O_2$.
This causes one to be out by at most just an automorphism of the final object, $O_{21}^{\sf\si\sn} - O_{12}^{\sf\si\sn} = Aut(O^{\sf\si\sn})$.}}
\label{Pentagon}\end{figure}            }

\section{Towers and digraphs of levels of mathematical structure}\label{TD}

{\bf Structure 1} Let us now view Figure 2's generalization to Fig 6.e) as what can be {\sl posed} at each level of mathematical structure. 
Since we have argued for universality of the claws, CoToBI should concentrate on what {\sl can} vary. 
The first such item is whether each level succeeds in realizing both, either, or none of the 2-way routes. 
Wheelerian Two-way routes thus constitute Selection Principles. 

\m 

\n{\bf Structure 2} The Constructabilities moreover transcend levels of structure. 
This is obvious for the single-primality Constructabilities: Space from less structured Space assumed and Spacetime from less structured Spacetime assumed. 
As regards the primality-traversing Spacetime from Space Constructability, this can occur a) within a common level: one of the 2-way routes. 
It can however also involve b) obtaining a more structured level's spacetime from a less structured level's space. 
a) and b) can moreover jointly occur, since Constructability can exhibit branching in the Brackets Closure Algorithm.   
Branching also means that multiple versions of b) can occur as well.  

\m 

\n The levels of structure can be viewed as a `multi-storey car park' of levels of mathematical structure. 
Thereby, some theories are found to involve further levels of structure than initially contemplated. 
This corresponds to there being `{\bf ramps}' between different levels in our car park.

\m 

\n While some simple models of equipping with additional mathematical structure involve a tower (alias chain) of levels, 
such equipping more generally can branch and loop, thus forming a digraph. 
CoTOBI involves posing Fig 6.e) at each level of structure within such a digraph, with Constructability ramps providing inter-relations between some of the levels. 
In the `Equipped Sets Foundational System of Mathematics' \cite{ASoS, ABook}, moreover, such digraphs sprout from the root of the bare (i.e.\ unequipped) sets.  
In the current Article, we stay within this most habitual Foundational System for simplicity.  

\m 

\n Let us next illustrate these phenomena with examples.  

\m 

\n{\bf Example 0} In flat space, translation-invariance and rotation-invariance can be entertained separately as well as together. 
This gives the 4-cycle digraph of Fig \ref{New-Gates-3}.a). 
This illustrates both branching and looping in the process of further equipping a set.  
Neither translations by themselves nor rotations by themselves imply the other \cite{FileR}, however, 
so this 4-cycle digraph of levels of structure does not contain any ramps.  

\m 

\n{\bf Example 1} Still working on flat space, suppose that one adjoins the general quadratic generator to the general inhomogeneous-linear generators 
that ab initio form the affine group $Aff(d)$; set also $d \geq 3$ to avoid unnecessary complications.  
Then branching occurs \cite{A-Brackets, IX}, via the presence of two strong factors in the Lie Algorithm in the schematic form of eq.\ (\ref{Branch}). 
Setting the first strong factor to zero amounts to restricting the general quadratic generator to the special-projective generator. 
Setting the second strong factor to zero amounts to restricting the general quadratic generator to the special-conformal generator.  
In this second case, moreover, consistency requires the affine group to be restricted to the similarity group, 
by shear and $d$-volume preserving squeeze refusing to close with the special-conformal generator. 
This gives the branching of levels of structure exhibited in Fig \ref{New-Gates-3}.b) (but no looping of ramps).   
(It should now be clear that the complications being avoided are infinitely-generated conformal groups in dimension $\leq 2$.)  

\m 

\n{\bf Example 2} Example 1)'s working carries over to the flat indefinite spacetime setting. 
This gives us an example of Spacetime Construction from less Spacetime structure Assumed, with likewise branching as well as just ramps.  

\m 

\n{\bf Example 3} General-coefficient geometrodynamics rigidly restricts to DeWitt coefficient geometrodynamics, i.e.\ GR in dynamical form.  
If done in the presence of minimally-coupled matter, universal local Lorentzian relativity also ensues.
This is an example of succeeding in realizing Constructability of Spacetime from Space within a given level of mathematical structure 
(metric Differential Geometry in its usual setting in which diffeomorphisms act).  

\m 

\n{\bf Example 4} A somewhat more general deformation \cite{AM13, ABook, IX} branches into the preceding, 
a Galileo--Riemann geometrostatics' space-time structure alongside universal Galilean   relativity, 
and strong-gravity geometrodynamics' spacetime structure alongside universal Carrollian relativity.
This example is of foundational interest, the first two branches being Einstein's Dilemma of the form to be taken by universal relativity, 
now arising as the roots of simple algebraic equations from Dirac's Algorithm. 
The third (Carrollian) branch, subsequently considered, corresponds to allowing for zero as well as finite or infinite universal propagation speed 
(commonly referred to as speed of light). 
Its somewhat limited, and yet physically significant, application is to Gravitational Theory's regime in the vicinity of spacetime singularities.  
It also serves as an example of branching occurring in the Dirac Algorithm, and of multiple different notions of spacetime (or space-time) structure 
arising from a single Dirac Algorithm calculation by means of such branching.  

\m 

\n{\bf Remark 1} This is suggestive that `omitting further study of branching Dirac Algorithms' \cite{HT92} 
is no longer a viable position in future treatises of Dirac Algorithms or generalizations thereof.  
Present indications are rather that, once Constructability is taken into account, 
branching is actually both common and a phenomenon with substantial geometrical and physical content. 

\m 

\n{\bf Example 5} Either of the preceding workings is accompanied by \cite{RWR, AM13, ABook, IX} a version involving a further weakly vanishing condition: 
constant mean curvature (CMC) slicing.  
This gives conformogeometrodynamics (at the level of Conformal Differential Geometry) from assumptions of just geometrodynamics (Differential Geometry).  
The most usual realization of this is as the most standard and successful way of approaching GR's Initial-Value Problem 
(finding data compatible with GR's constraints). 
This was originally conceived of by noting that CMC slicing decouples GR's constraints \cite{York72}. 
CMC has however since also been found to confer consistency to deformed versions of GR's constraints.
This constitutes a second foundation pointing to the involvement of CMC alongside gravitational constraints, now arising as a weak branch from the Dirac Algorithm. 

\m 

\n{\bf Example 6} Assuming just metrostatics -- spatial metrics without spatial diffeomorphism irrelevance -- 
along the GR-like branch, one runs into inconsistency \cite{OM02-San} unless the spatial diffeomorphism encoding momentum constraint $\u{\scM}$ is included anyway. 
This is because it is discovered as an integrability of the GR-like Hamiltonian constraint $\scH$ (a fact first discovered by Moncrief and Teitelboim \cite{MT72}). 
So in this case there is no Constructability of Spacetime from Space at the level of metrics not modulo diffeomorphisms. 
There is however \cite{AM13} a Constructability of spacetime metrics modulo diffeomorphisms from spatial metrics without diffeomorphisms. 
I.e.\ half of the 2-way route exists solely in lifted ramp form for this example.  
Note that this example includes a 3-cycle loop forming the digraph in Fig \ref{New-Gates-3}.c). 
In contrast to Example 3's loop, however, this one has a ramp along each edge. 
This illustrates that ramps can loop, in this case by having a branch that permits one to 'miss a floor', 
here going straight from metrodynamics to conformogeometrodynamics without passing through geometrodynamcis.  
%
{            \begin{figure}[!ht]
\centering
\includegraphics[width=0.85\textwidth]{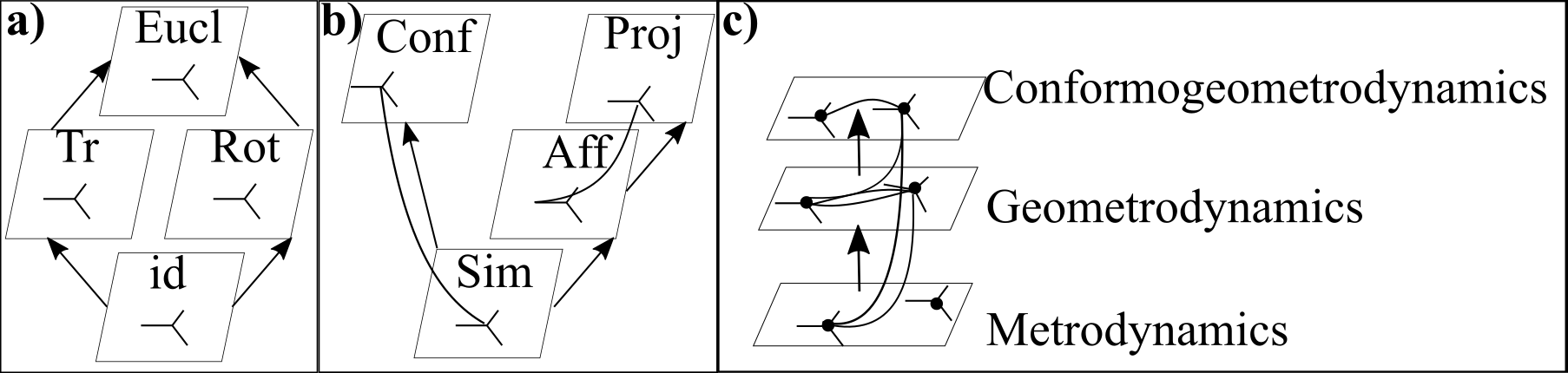} 
\caption[Text der im Bilderverzeichnis auftaucht]{\footnotesize{Levels of structure, and ramps therebetween, for each of Examples 0 to 6.} } 
\label{New-Gates-3} \end{figure}          }

\m 

\n The information of whether each level of mathematical structure's 2-way route works is characterized by the following. 

\m 

\n{\bf Definition 1} The {\it Wheeler matrix} is the binary-valued matrix encoding which of the $f$ floors under consideration contain routes. 

\m 

\n{\bf Remark 2} In Physics, it is a $2 \times f$ matrix, whereas in pure Geometry, it is absent.  

\m 

\n{\bf Definition 2} A {\it Wheelerian floor} has entries (1, 1) to indicate a 2-way route.  
An {\it anti-Wheelerian floor} has entries (0, 0) to indicate no routes. 

\m 

\n{\bf Remark 3} It is also possible to just have a forward route or just a backward route, as encoded by (1, 0) and (0, 1) floors in the Wheeler matrix. 
Within a tower of levels, the Wheeler matrix would be ordered along the chain of structure. 
Within a more general digraph of levels, it could be ordered more arbitrarily.
Or, if the shape of the digraph is considered to be sufficiently important, one could use the following instead. 

\m 

\n{\bf Definition 3} A {\it Wheeler labelled digraph} is the levels of mathematical structure digraph of relevance to one's problem in hand. 
Now with each vertex (representing a level) labelled with its 2-bit entry of Wheeler data concerning how much of that level's 2-way route exists. 

\m 

\n{\bf Definition 4} The {\it adjacent ramp matrix} is the binary-valued matrix of ramps between adjacent levels in a tower. 
This has a 1 entry when a ramp exists and a 0 entry when not.  

\m 

\n{\bf Remark 4} It is    a  $3 \times (f - 1)$ matrix in Relativistic Physics 
                 but just an          $(f - 1)$-vector for pure Geometry. 
 
\m 

\n{\bf Definition 5} Within a more general digraph of levels, the {\it adjacent ramp matrix} has $3 \times e(L)$ entries, for edge number $e(L)$ \cite{Graph-2}.  

\m 

\n{\bf Remark 5} One can also consider how many levels one can jump in one go by considering the {\it ramp digraph} with $3 \times c(L)$ entries, 
for $c(L)$ the {\it chain number}: total number of chains  within $L$.  

\m 

\n{\bf Remark 6} It is of course possible to return to Linear Algebra by using {\it adjacency matrices for ramp digraphs}. 
(Adjacency matrices are a standard piece of basic Graph Theory \cite{Graph-1, Graph-2}, not to be confused with ramps solely between adjacent levels of structure).
%
{            \begin{figure}[!ht]
\centering
\includegraphics[width=0.85\textwidth]{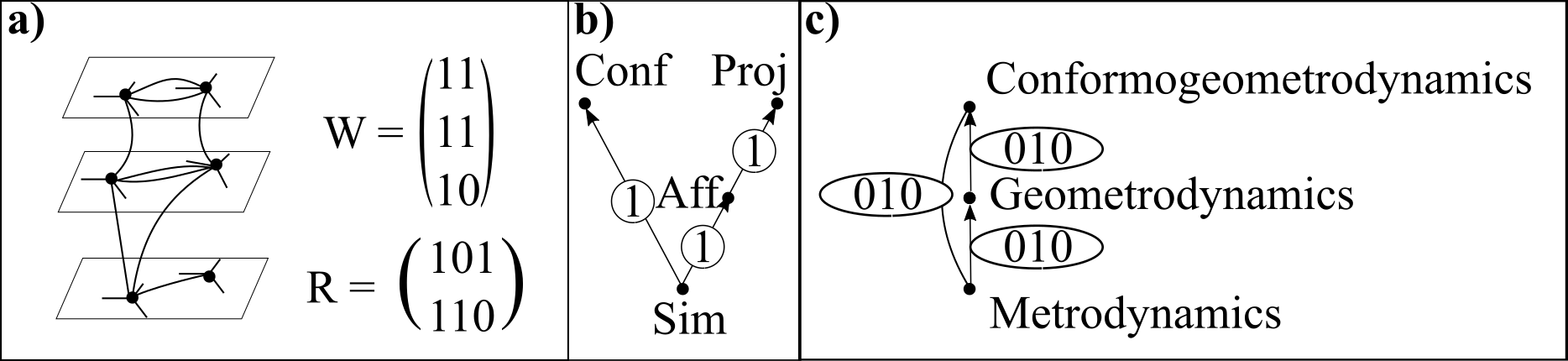} 
\caption[Text der im Bilderverzeichnis auftaucht]{\footnotesize{a) Wheeler and ramp matrices for some arbitrary tower. 
b) Ramp labelled digraph for Flat Geometry, exhibiting branching.
c) Ramp labelled digraph for (conformo)(geo)metrodynamics.} } 
\label{Tower-2} \end{figure}          }

\section{Conclusion}\label{Conclusion}

\n We extended consideration of Background Independence from the habitual Differential Geometry setting for physical laws 
to the general level of mathematical structure (within the standard Equipped Sets Foundational System of Mathematics). 
The Differential Geometry level version largely revolves around the Lie claw digraph of Background Independence aspects, 
with Closure at its centre interlinked separately with Relationalism, Observables and Constructability on its periphery.  
In the current Article, we argued that these four aspects are already categorical, 
by which such claws are in fact universal: present at any level of mathematical structure. 
Because of this, we name them in general {\sl Universal Background Independence Claws (UBIC)}.  
(That this concerns (resolutions of) a number of major and longstanding fundamental issues in Quantum Gravity and the nature of Physical Law 
is clear from tracking these aspects' names and notions back to the list of foundational problems in Appendix A. 
For instance, the Lie claw digraph and the below Wheelerian two-way route gives A Local Resolution of the Problem of Time (ALRoPoT) at the classical level 
\cite{ABook, ALett, XIV, I, II, III, IV, V, VI, VII, VIII, IX}.  
The current Article's categorical generalization serves moreover to partly resolve, and elsewise sharply pose quantum-level counterparts.)

\m 

\n Our approach principally concerns automorphisms and bracket algebraic structures being both categorically meaningful 
and the natural mathematics of the centre of the claw.
The current Article introduces the General Brackets Closure Algorithm extension of the generalized Lie Algorithm \cite{XIV}: 
now for general, rather than just Lie, brackets algebraic structures.
(The generalized Lie Algorithm is itself an extension of Lie's Algorithm to include insights of Dirac's 
that were previously only considered in the much narrower context of Poisson brackets of constraints).\footnote{This generality comes with the following 
technical caveat. 
While the Nambu, Nijenhuis and Gerstenhaber generalizations of Lie are known \cite{Nambu, Nijenhuis} to provide the complete claw structure, 
Loday's distinct generalization of note \cite{KS} remains unchecked in one detail (whether Observables still decouple from Closure-and-Relationalism). 
It is however not necessarily clear whether even {\sl Quantum} Gravity will require use of Loday brackets 
(Nijenhuis and Gerstenhaber being at least prototypical for quantum operator algebras, and Nambu for M-Theory).}

\m 

\n Universality means that there is one claw per level  of mathematical structure in Geometry, 
or two claws per level of mathematical structure in Physics (spacetime primality copy and space/dynamics/canonical copy). 
So in the Comparative Theory of Background Independence (CoToBI), 
one is therefore to concentrate on which {\sl other} features Background Independence possesses that {\sl can} vary from level to level. 
Such features thus provide Selection Principles for how far along the levels of mathematical structure Background Independence can be entertained 
(and ALRoPoT extended to).   

\m 

\n Such features are, firstly, the Wheelerian two-way routes between the above two copies at a given level. 
Secondly, Constructability from less structure assumed provides `ramps' linking between some pairs of levels of mathematical structures, 
in the manner of a multi-storey car park.  

\m 

\n Constructabilities are is set up by Deformation of Generators and succeeds when Rigidity is encountered.
This is in turn underlied by cohomological triviality conditions (which {\sl can} vary from category to category as to whether they are realized).  

\m 

\n In fact, one of the two Wheelerian routes is itself a Constructability: of Spacetime from Space \cite{RWR, AM13, ABook, IX}, 
and is capable of also being a ramp: obtaining spacetime from {\sl a less structured level's notion of} space.  
(This capacity is also underlied by Brackets Closure Algorithms' capacity to branch, by producing strings of cofactors, each of which vanishing may 
lead to a different consistent theory.  
By this, one can have Spacetime from Space provide both an intra-level Wheelerian route and one or more inter-level ramps, of which Sec 4 provided an example.
That branching can thus no longer \cite{HT92} be ignored in study of Brackets Closure Algorithms is itself a significant conclusion reached by the current Article.)

\m 

\n The other Wheelerian route is the Refoliation Invariance resolution of Foliation Independence in GR, 
generalizing to Reslicing Invariance resolution of Slicing Independence more generally. 
Algebraically, this is a Reallocation of Intermediary Object (RIO) Invariance: a commuting pentagon involving assigning intermediary objects in either 
order as its four side edges, whose top edge closes if this causes difference by at most an automorphism of the final object.  
The point is then that, while Constructabilities and RIO Invariance can be {\sl posed} for arbitrary levels of structure, 
their affirmative resolution is {\sl not} categorical, and so is available to serve as a Selection Principle.  

\m 

\n We further characterize these key Selection Principle features by introducing {\it Wheeler} and `{\it ramp matrices}.  
Since equipping levels of mathematical structure can in fact branch and loop (Section 4 providing examples), 
one more generally requires {\it Wheeler} and {\it ramp labelled digraphs}.   
At least at the local level, CoTOBI would appear to be largely shaped by these matrices or labelled digraphs. 

\m 

\n There are moreover some indications that CoToBI is largely a global subject 
(\cite{A-CBI} but also the above-mentioned deformed algebraic structure cohomology conditions).  
UBIC, Wheelerian routes and ramps tell us moreover {\sl which} structures to globalize, and to consider quantum versions of. 
In particular, via its categorical generality, UBIC bodes well as regards the quantum counterpart of the current work being meaningfully posed.  

\m 

\n{\bf Acknowledgments} I thank Professor Chris Isham for discussions over the past decade, 
                                          Przemek Malkiewicz for discussions over the past five years, 
and as yet unnamed people for more recent conceptual, topological and combinatorial discussions.						
Professors Malcolm MacCallum, Don Page, Reza Tavakol and Enrique Alvarez for career support, and various close people for support.    

\begin{appendix}

\section{From Problem of Time aspects to categorically meaningful Background Independence aspects}\label{App}
%
This Appendix serves to recognize previous names and notions of Problem of Time facets and Background Independence aspects, 
as well as to connect from previous such to their current forms. 
For their roots in more basic spatio-temporal properties, and predecessors of BI in the work of Leibniz, Mach and Einstein, see Part I of \cite{ABook}. 
%
{            \begin{figure}[!ht]
\centering
\includegraphics[width=1.0\textwidth]{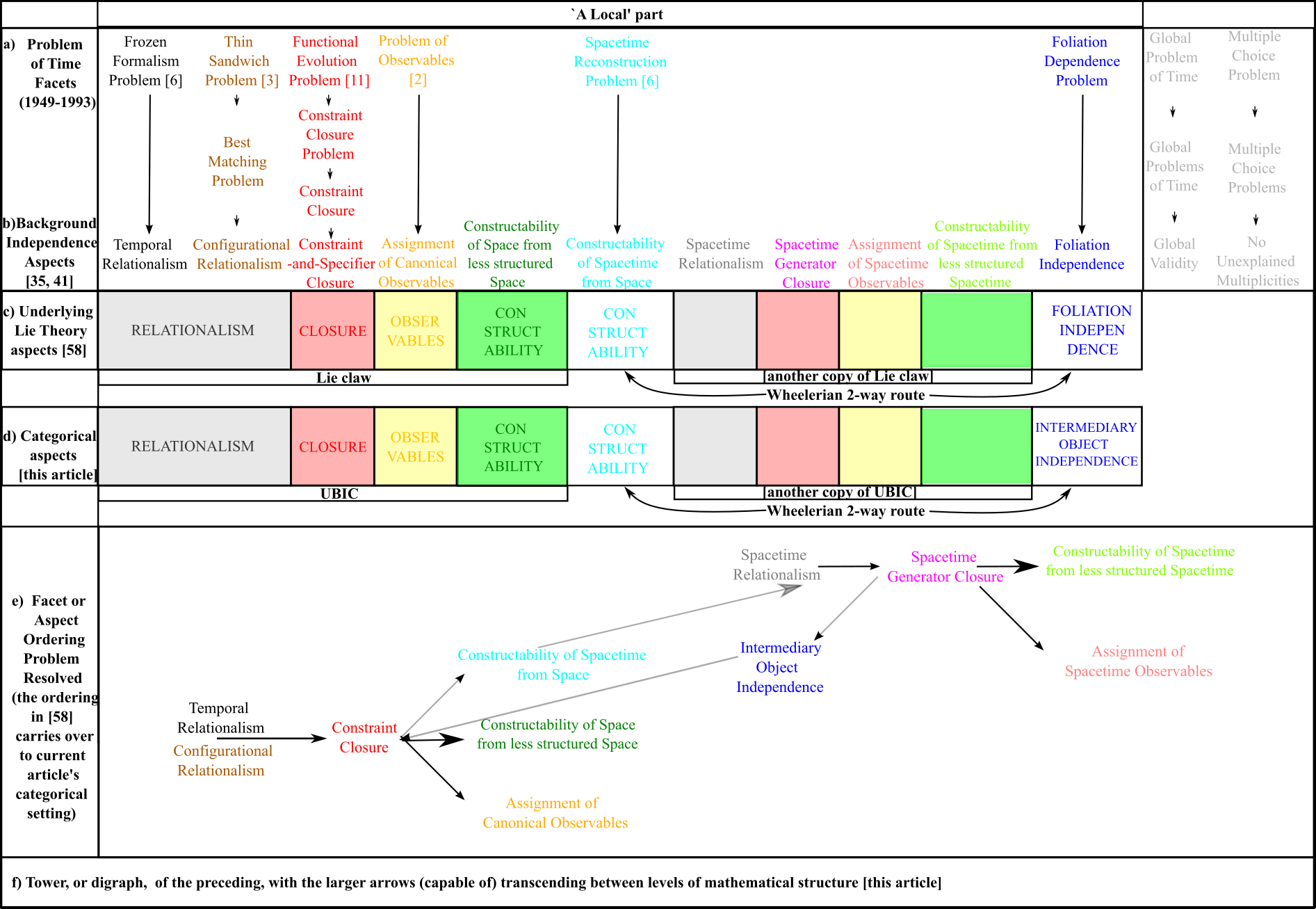} 
\caption[Text der im Bilderverzeichnis auftaucht]{\footnotesize{Evolution from Problem of Time facets a) 
to categorically meaningful Background Independence aspects d), e), f).} } 
\label{UBIC-App-Fig} \end{figure}          }

\end{appendix}


\end{document}